\documentclass[usenatbib,useAMS]{mnras}

\usepackage{graphics,epsf}
\usepackage{amsmath}                
\usepackage{amsfonts}               
\usepackage{amssymb}                
\usepackage{epsfig}                 
\usepackage{graphicx}

\usepackage{xcolor}
\definecolor{redak}{rgb}{0.9,0.15,0.05}

\def \kms{~\rm{km~s^{-1}}}

\def \msyr{~\rm{M_{\odot}}~\rm{yr^{-1}}}

\def \K{~\rm{K}}

\def \au{~\rm{au}}

\def \etc{$\eta$~Car~}
\def \days{~\rm{days}}

\def \rmModot{~\rm{M_{\sun}}}
\def \rmRodot{~\rm{R_{\sun}}}

\defcitealias{Kashi2017}{Paper~I}

\title[Accretion Response Simulations of Eta Carinae]{Simulating the response of the secondary star of Eta~Carinae to mass accretion at periastron passage}

\author[A. Kashi]{Amit Kashi$^{1}$\thanks{E-mail: \href{mailto:kashi@ariel.ac.il}{kashi@ariel.ac.il}} \\
$^{1}$Department of Physics, Ariel University, Ariel, POB 3, 40700, Israel\\
}

\date{Accepted 2019 March 19. Received 2019 March 19; in original form 2018 May 05}

\pubyear{2019}

\begin{document}
\label{firstpage}
\pagerange{\pageref{firstpage}--\pageref{lastpage}}
\maketitle

\begin{abstract}
We use high resolution 3D hydrodynamical simulations to quantify the amount of mass accreted onto the secondary star of the binary system Eta Carinae, exploring two sets of stellar masses that had been proposed for the system, the conventional mass model ($M_1=120 \rmModot$ and $M_2=30 \rmModot$) and the high mass model ($M_1=170 \rmModot$ and $M_2=80 \rmModot$).
The system consists of two very massive stars in a highly eccentric orbit. Every cycle close to periastron passage the system experiences a spectroscopic event during which many lines change their appearance, accompanied by a decline in x-ray emission associated with the destruction wind collision structure and accretion of the primary wind onto the secondary.
We take four different numerical approaches to simulate the response of the secondary wind to accretion. Each affects the mass loss rate of the secondary differently, and in turn determines the amount of accreted mass.
The high mass model gives for most approaches much more accreted gas and a longer accretion phase.
We find that the effective temperature of the secondary can be significantly reduced due to accretion.
We also test different eccentricity values and a higher primary mass loss rate and find their effect on the duration of the spectroscopic event.
We conclude that the high mass model is better compatible with the amount of accreted mass, $\approx 3 \times 10^{-6} \rmModot$, required for
explaining the reduction in secondary ionization photons during the spectroscopic event and compatible with its observed duration.
\end{abstract}


\section{INTRODUCTION}
\label{sec:intro}

The binary system $\eta$ Carinae is composed of a very massive star at late stages of its evolution, the primary, and a hotter and less luminous evolved main sequence star, the secondary  (\citealt{Damineli1996, DavidsonHumphreys1997,DavidsonHumphreys2012}).
The binary system has a highly eccentric orbit (e.g., \citealt{Daminelietal1997, Smithetal2004, Davidsonetal2017}), and strong winds (\citealt{PittardCorcoran2002, Akashietal2006}) resulting in unique period of strong interaction every 5.54 years during periastron passage known as the spectroscopic event.
During the event many spectral lines and emission in basically all wavelengths show rapid variability (e.g.
\citealt{Zanellaetal1984, Davidsonetal2000, Smithetal2000, DuncanWhite2003, Whitelocketal2004, Martinetal2006, Martinetal2010, Stahletal2005, Hamaguchietal2007, Hamaguchietal2016, Nielsenetal2007, Daminelietal2008a, Daminelietal2008b, Mehneretal2010, Mehneretal2011, Mehneretal2015, Davidson2012}).
The x-ray intensity, which also serves as an indicator to the intensity of wind interaction, drops for a duration of a few weeks, changing from one spectroscopic event to the other (\citealt{Corcoran2005, Corcoranetal2010, Corcoranetal2015} and references therein).

\cite{Soker2005b} developed a model to interpret the line variations during the spectroscopic event as a result of accreting clumps of gas onto the secondary near periastron passages, disabling its wind. The suggestion was later developed to a detailed model accounting for different observations in the accretion model framework (\citealt{Akashietal2006}; \citealt{KashiSoker2009a}).

The last three spectroscopic events, 2003.5, 2009 and 2014.6 were not similar, and reflected a trend in the intensities of various lines \citep{Mehneretal2015}.
Observations of spectral lines across the 2014.6 event can be interpreted as weaker accretion onto the secondary close to periastron passage compared to previous events.
This may indicate a decrease in the mass-loss rate from the primary star claimed to be a `change of state', already identified by \cite{Davidsonetal2005}, and theoretically explained by \cite{Kashietal2016}.
Further indication for the change of state were recently found from comparison of UV lines emission at similar orbital phases separated by two orbital revolutions, at positions far from periastron passage \citep{Davidsonetal2018}.

\cite{KashiSoker2009b} performed a more detailed calculation, integrating over time and volume of the density within the Bondi-Hoyle-Lyttleton accretion radius around the secondary,
and found that accretion should take place close to periastron and the secondary should accrete $\sim~\rm{few}~\times~10^{-6}~\rmModot$ each cycle.

Older grid-based simulations \citep{Parkinetal2011} and SPH simulations \citep{Okazakietal2008, Maduraetal2013} of the colliding winds have not obtained accretion onto the secondary. \cite{Teodoroetal2012} and \cite{Maduraetal2013} advocated against the need of accretion in explaining the spectroscopic event.

\cite{Akashietal2013} performed 3D numerical simulations using the \texttt{VH-1} hydrodynamical code \citep{Blondin1994,Hawleyetal2012} to study the accretion and found that a few days before periastron passage clumps of gas are formed due to instabilities in the colliding winds structure, and some of these clumps flow towards the secondary implying accretion should occur.

The final theoretical evidence for accretion came from simulations in \citet[hereafter \citetalias{Kashi2017}]{Kashi2017}.
These simulations showed the destruction of the colliding winds structure into filaments and clumps that later flew onto the secondary.
\citetalias{Kashi2017} demonstrated that dense clumps are crucial to the onset of the accretion process.
The clumps were formed by the smooth colliding stellar winds that developed instabilities that later grew into clumps (no artificial clumps were seeded).
This confirmed preceding theoretical arguments by \cite{Soker2005a,Soker2005b} that suggested accretion of clumps.
The amount of accreted mass was not derived from the simulations in \citetalias{Kashi2017}, as it required further modeling of the secondary star response to the accreted mass.

It is expected that accretion will cause the secondary star to stop, or partially stop, blowing its wind (\citealt{KashiSoker2009b} and ref. therein).
However, quantifying the effect is a complicated task.
One should consider the acceleration mechanism of the wind (line driving in the case of the secondary), and how gas settling on the envelope will reduce it.
As gas is coming both from filaments and clumps directly,
the wind of the secondary is expected to be affected directionally rather than isotropically.

In this work we take a step forward, and quantify the accretion process and its dependence on the different parameters.
In section \ref{sec:simulation} we describe the numerical simulation.
Our results, showing accretion quantitatively, are presented in section \ref{sec:results}.
Our a summary and discussion in given in section \ref{sec:summary}.

\section{THE NUMERICAL SIMULATIONS}
 \label{sec:simulation}

We use version 4.5 of the hydrodynamic code \texttt{FLASH}, originally described by \cite{Fryxell2000}.
Our 3D Cartesian grid extends over $(x,y,z) = \pm 8 \au$, centered around the secondary.
Our initial conditions are set $50$ days before periastron, which is enough time for forming the colliding winds structure (also know as the wind-wind collision zone, or WWC).
We place the secondary in the center of the grid and send the primary on an eccentric $e\simeq0.85$--$0.9$ Keplerian orbit.
As the mass loss and mass transfer during present-day \etc are small (in contrast to their values during the GE and LE), the deviation from Keplerian orbit is very small.
As the simulations are performed in the secondary rest frame, the wind is ejected isotropically around the secondary, and non-isotropically around the primary, as its orbital velocity around the secondary is subtracted from the wind velocity.
To solve the hydrodynamic equations we use the \texttt{FLASH} version of the split piecewise parabolic method (PPM) solver \citep{ColellaWoodward1984}.
We use the code's ratiation-transfer multigroup diffusion approximation, with one energy group (similar to the Eddington gray approximation).
We use five levels of refinement with better resolution closer to the center. The length of the smallest cell is $1.18 \times 10^{11} \rm{cm}$ ($\simeq 1.7\rmRodot$). 
This finest resolution extends over a sphere of a radius of $\simeq 82 \rmRodot$ centered at $(0,0,0)$.
The second finest level of resolution, resolving twice the spatial scale of the finest level, continues up to a radius of $\simeq 320 \rmRodot$.
This level of resolution covers the apex of the colliding winds from $\simeq 20 \days$ before periastron and on.
As shown below and discussed in \citetalias{Kashi2017}, the instabilities that lead to accretion start only a few days before periastron, namely within this level of resolution.
The highest resolution allows to follow in great detail the gas as it reaches the injection zone of the secondary wind and being accreted onto the secondary.

\begin{figure}
\centering
\includegraphics[trim= -0.5cm -0.5cm 0.0cm 0.0cm,clip=true,width=0.99\columnwidth]{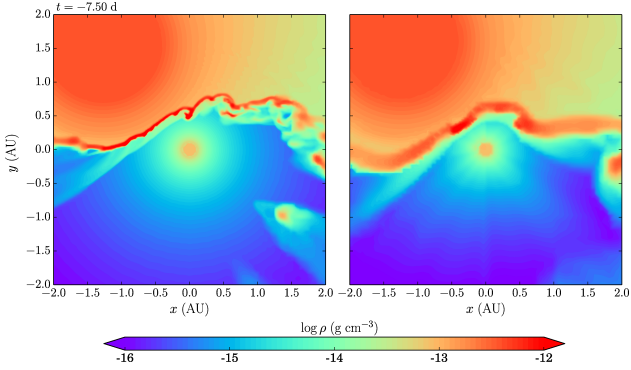}  
\caption{
Density maps showing the density sliced in the orbital plane ($z=0$), for the conventional mass model ($M_1=120 \rmModot$ and $M_2=30 \rmModot$) at two resolutions.
The secondary is at the center, and the primary orbits it from the upper part of the figure to the bottom-left until periastron, and then down-right.
Periastron occurs at $(x,y,z)=(-1.664 \au, 0, 0)$ and $t=0$.
The two panels show the stars and the colliding wind structure 7.5 days before periastron.
The left panel shows the simulation results with the high resolution described in section \ref{sec:simulation}, while the right panel shows a simulation with lower resolution by 2 levels of refinement, namely the spatial scale resolved is 4 times larger.
}
\label{fig:resolution_comparison}
\end{figure}

Figure~\ref{fig:resolution_comparison} shows the same simulation at two different resolutions, manifesting the higher details revealed by the higher resolution.
The right panel shows a simulation with lower resolution by 2 levels of refinement, namely the spatial scale resolved is 4 times larger.
It can be seen that the high resolution simulation:

\noindent(1) prevents unwanted effects of the grid that makes deviations from spherical symmetry, as can be seen in the right panel of Figure~\ref{fig:resolution_comparison} that shows that the density of the secondary wind is not perfectly isotropic.

\noindent(2) much better resolves the secondary as a sphere, which is important for resolving directional accretion.

\noindent(3) much better resolves the colliding winds structure with the two shocks and a contact instability.

\noindent(4) allows small scale instabilities to form, which consequently create filaments and clumps, some of which later get accreted by the secondary.

We therefore conclude that the high resolution is absolutely essential for obtaining meaningful results from the simulation. Therefore we ran all our simulations at the high resolution described above. 
%

In \citetalias{Kashi2017} we took into account self-gravity. 
However, we found that the formation of filaments and clumps that are later accreted onto the secondary can occur as a result of instabilities that do not involve self-gravity.
The free-fall (collapse) time of each clump as a result of self-gravity is much longer than the duration of the clump formation, indicating that self-gravity does not have
a significant role in the formation of the clumps.
We therefore here disable self gravity and consider only the gravity of the two stars, modeled as point masses.

As there are different arguments in the literature regarding the masses of the two stars, we use two sets of stellar masses, similar to the sets we used in \citetalias{Kashi2017}:
\begin{enumerate}
\item
\emph{Conventional mass model}, where the primary and secondary masses are $M_1=120 \rmModot$ and $M_2=30 \rmModot$, respectively \citep{Hillieretal2001}.
\item \emph{High mass model} with $M_1=170 \rmModot$ and $M_2=80 \rmModot$ (\citealt{KashiSoker2010}, where the model was referred to as the `MTz model'; \citealt{KashiSoker2015}).
\end{enumerate}

The orbital period is $P=2023$ days, implying the semi-major axis is $a=16.64 \au$ for the conventional mass model (e.g., \citealt{Ishibashietal1999, Daminelietal2000, Whitelocketal2004, Davidsonetal2005} and references therein), and $a=19.73 \au$ for the high mass model.
The stellar radii are taken to be $R_1=180 \rmModot$ and $R_2=20 \rmRodot$ for both the conventional and the high mass models (see \citealt{KashiSoker2008} for a discussion on how the radii are derived).
Larger secondary radius would probably make accretion somewhat easier.
The mass loss rates and wind velocities are $\dot{M}_1=3$--$10 \times 10^{-4} \msyr$, $v_1=500 \kms$
and $\dot{M}_2=10^{-5} \msyr$, $v_2=3\,000 \kms$, for the primary and the secondary, respectively \citep{PittardCorcoran2002, Hamaguchietal2007, Grohetal2012, Maduraetal2013, Corcoranetal2015}.
The range for $\dot{M}_1$ is due to both uncertainty in the value itself, and a possible decrease it is undergoing in the last $\approx15$ years, as mentioned in section \ref{sec:intro}.
The wind is being injected radially at its terminal speed from a narrow sphere around each star (see \citetalias{Kashi2017} for further details).
In the process of injecting the winds we neglect the spins of the stars, but the orbital motion of the primary relative to the fixed grid is taken into account. The accelaration of the winds is also neglected and both winds are ejected at their terminal speeds.
For the winds the adiabatic index is set to $\gamma=5/3$.
Our initial conditions at $t=-50 \days$ set the entire grid (except the stars themselves) filled with the smooth undisturbed primary wind.
We include radiative cooling based on solar composition from \cite{SutherlandDopita1993} according to the implementation described and tested in \citetalias{Kashi2017}.
\begin{table}
\centering
\caption{List of stellar and orbital parameters.
}
\begin{tabular}{lll} 
\hline
Parameter      & Conventional                       & High Mass         \\
               & model                              & model             \\
\hline
$M_1$          & $120\rmModot$                      & $170\rmModot$     \\
$M_2$          & $30\rmModot$                       & $80\rmModot$      \\

$R_1$          & \multicolumn{2}{l}{\qquad $180\rmRodot$}                      \\
$R_2$          & \multicolumn{2}{l}{\qquad $20\rmRodot$}                       \\

$v_1$          & \multicolumn{2}{l}{\qquad $500 \kms$}                         \\
$v_2$          & \multicolumn{2}{l}{\qquad $3\,000 \kms$}                      \\

$\dot{M}_1$    & \multicolumn{2}{l}{\qquad $3$--$10 \times 10^{-4} \msyr$}     \\
$\dot{M}_2$    & \multicolumn{2}{l}{\qquad $10^{-5} \msyr$}                    \\

$P$            & \multicolumn{2}{l}{\qquad $2023$ days}                        \\
$a$            & \multicolumn{2}{l}{\qquad $16.64 \au$}                        \\
$e$            & \multicolumn{2}{l}{\qquad $0.85$--$0.9$}                      \\

\hline
&&\\
\end{tabular}
\label{table:stellarandorbitalparameters}
\end{table}

As to the response to the secondary star to the accreted gas we take four approaches:

\noindent(1) Approaching gas removal: In the first approach we remove dense gas that reaches the secondary wind injection region, and replace it by fresh secondary wind with its regular mass loss and velocity. Namely, we do not make any changes to secondary wind and let it continue to blow as if the accreted gas did not cause any disturbance.

\noindent(2) Exponentially reduced mass loss: In the second approach we reduce the mass loss rate of the secondary as it approaches periastron passage according to
\begin{equation}
\dot{M}_{2, \rm{eff}}=
\left\{
\begin{array}{lc}
      \dot{M}_2                    & t\leq -5 ~\rm{d} \\
      \dot{M}_2 \exp[{-(t+5~\rm{d})\ln{10}/5~\rm{d}}] & -5 ~\rm{d} < t\leq 0 ~\rm{d}\\
      0.1\dot{M}_2                       & 0 ~\rm{d} < t \\
\end{array} 
\right. ,
\label{eq:mdot_approach2}
\end{equation}
where $t=0$ is the time where the system is at periastron passage.
This is an artificial approach that does not relate to the actual accretion situation in the simulation.
Also, note that $\dot{M}_2$ is kept on the low value of the remainder of the simulation. Namely, for this approach we do not apply recovery from accretion.

\noindent(3) Accretion dependent mass loss: In the third approach we dynamically change the mass loss rate of the secondary wind in response to the mass that has been accreted.
We lower $\dot{M}_2$ by changing the density of the ejected wind by the \emph{extra}
density of the accreted gas, namely
\begin{equation}
\begin{split}
\frac{d\dot{M}_{2, \rm{eff}}}{d\Omega} &=
\frac{d\dot{M}_2}{d\Omega} \frac{\rho_u(\Omega)-[\rho(\Omega)-\rho_u(\Omega)]}{\rho_u(\Omega)} \\
&= \frac{d\dot{M}_2}{d\Omega} \left(2-\frac{\rho(\Omega)}{\rho_u(\Omega)} \right).
\end{split}
\label{eq:mdot_approach3}
\end{equation}
In the above equation $\Omega$ is a solid angle, $d\dot{M}_2/d\Omega$ is the differential mass loss of the secondary, $\rho_u(\Omega)$ is the undisturbed density of the secondary wind as if it blows without the interruption of accreted gas, and $\rho(\Omega)$ is the density of the gas (if no accreted gas arrives to the secondary wind ejection region into a solid angle $\Omega$, then practically $\rho(\Omega)=\rho_u(\Omega)$ for that solid angle).
Note that the approach gives mass loss rate that is not isotropic but rather dependent on the direction from which parcels of the accreted gas arrived.
This approach therefore can only be implemented if the secondary and its immediate vicinity (from where its wind is being ejected and to where mass from the primary wind is being accreted) are simulated with high resolution.

\noindent(4) No intervention:
In the fourth approach we do not remove any accreted gas from the simulation. Cells in the secondary wind injection zone where dense blobs arrive are not replaced by fresh secondary wind but rather kept as is, i.e., with the density, velocity, and temperature of the blob. If the blob reaches the innermost injection zone, then the mass-loss rate over the solid angle of that cell for that timestep is zero; otherwise, the mass-loss rate per solid angle is unchanged.

Table \ref{table:parameters} summarizes the simulations we ran, with different stellar masses and different approaches for treating the response of the secondary wind to the accreted gas. We also test denser primary wind and another value of orbital eccentricity.
\begin{table*}
\centering
\caption{List of simulations.
Run naming code:
C$=$ Conventional; M$=$ Massive; WA$=$ Wind Acceleration.
}
\begin{tabular}{lcccclcc} 
\hline
Run       & Stellar masses      &Semi-major     &Orbital      &$\dot{M}_1 (10^{-4} $       & Approach for secondary  & Accretion       & Accreted mass        \\
          & $M_1,M_2(\rmModot)$ &axis ($\au$)   &eccentricity &$\msyr$)                    & response to accretion   & duration (days) & $(10^{-6} \rmModot)$ \\
\hline
C1        & 120,30  &16.64  & 0.9  &6           & (1) Approaching gas          & 2         & 0.01             \\
          &         &       &      &            & \quad \enskip removal        &           &                  \\
C2        & 120,30  &16.64  & 0.9  &6           & (2) Exponentially reduced    & 65        & 3.8              \\
          &         &       &      &            & \quad \enskip mass loss      &           &                  \\
C3        & 120,30  &16.64  & 0.9  &6           & (3) Accretion dependent      & 2$^{a}$   & 0.04             \\
          &         &       &      &            & \quad \enskip mass loss      &           &                  \\
C4        & 120,30  &16.64  & 0.9  &6           & (4) No intervention                  & 3$^{b}$   & 0.2      \\
&&&&&&\\
&&&&&&\\
M1        & 170,80  &19.73  & 0.9  &6           & (1) Approaching gas          & 16        & 0.04             \\
          &         &       &      &            & \quad \enskip removal        &           &                  \\
M2        & 170,80  &19.73  & 0.9  &6           & (2) Exponentially reduced    & 65        & 4.2              \\
          &         &       &      &            & \quad \enskip mass loss      &           &                  \\
M3        & 170,80  &19.73  & 0.9  &6           & (3) Accretion dependent      & 29        & 0.06             \\
          &         &       &      &            & \quad \enskip mass loss      &           &                  \\
M4        & 170,80  &19.73  & 0.9  &6           & (4) No intervention          & 38        & 1                \\
&&&&&&\\
&&&&&&\\
C5        & 120,30 &16.64   & 0.9  &10          & (4) No intervention          & 3$^{c}$   & 0.2              \\
M5        & 170,80 &19.73   & 0.9  &10          & (4) No intervention          & 45        & 3.1              \\
&&&&&&\\
&&&&&&\\
C6        & 120,30 &16.64   & 0.85 &6           & (4) No intervention          & 3         & 0.4              \\
M6        & 170,80 &19.73   & 0.85 &6           & (4) No intervention          & 64        & 1.1              \\
&&&&&&\\
&&&&&&\\
M4WA$^{d}$& 170,80 &19.73   & 0.9  &6           & (4) No intervention          & 48        & 1.6              \\
\hline
&&\\
\end{tabular}
\begin{flushleft}
$^{a}$ Run C3 also shown long lasting very weak accretion, but the main accretion phase last $\simeq2$ days.
\newline
$^{b}$ Run C4 has 2 accretion episodes of clumps separated by many days.
\newline
$^{c}$ Run C5 has 3 accretion episodes of clumps separated by many days.
\newline
$^{d}$ Run M4WA is similar to run M4 in all parameters but includes wind acceleration for the secondary star.
\end{flushleft}
\label{table:parameters}
\end{table*}

Let us elaborate on how we calculate the accreted mass.
With no accretion, the volume around the secondary, in the injection zone of the wind is supposed to have a certain density in each cell according to
\begin{equation}
\rho_u(r)=\frac{\dot{M}_2}{4 \pi r^2 v_2} .
\label{eq:rhou}
\end{equation}
As the simulation runs, high density clumps and filaments approach the injection zone of the secondary wind and even reach the cells of the secondary itself. Whenever the actual density $\rho_{a, \rm{cell}}$ in a cell in the injection zone increases above the expected undisturbed value of $\rho_{u, \rm{cell}}$,
we count the extra mass as accreted
\begin{equation}
\Delta M_{\rm acc} = (\rho_{a, \rm{cell}} - \rho_{u, \rm{cell}}) V_{\rm{cell}},
\label{eq:Macc}
\end{equation}
where $V_{\rm{cell}}$ is the volume of the cell.
We then sum all the contributions from all cells in the injection zone to obtain the total mass accreted for that time step.

\section{RESULTS}
\label{sec:results}

We post-processed every simulation to measure the mass accreted onto the secondary and derive other quantities that we discuss below.
As the simulation is of very high resolution both the running time and the post-processing are long.
We therefore derive a post-processing output every $\simeq1/2$ day,
even though our data is calculated in time steps of $\simeq 1$--$3$ minutes (a necessary short time step determined by the Courant condition).
This interval is however sufficient to produce accretion rate and other quantities in a good accuracy.

\begin{figure}
\centering
\includegraphics[trim= 0.cm 0.cm 0.cm 0.cm,clip=true,width=0.99\columnwidth]{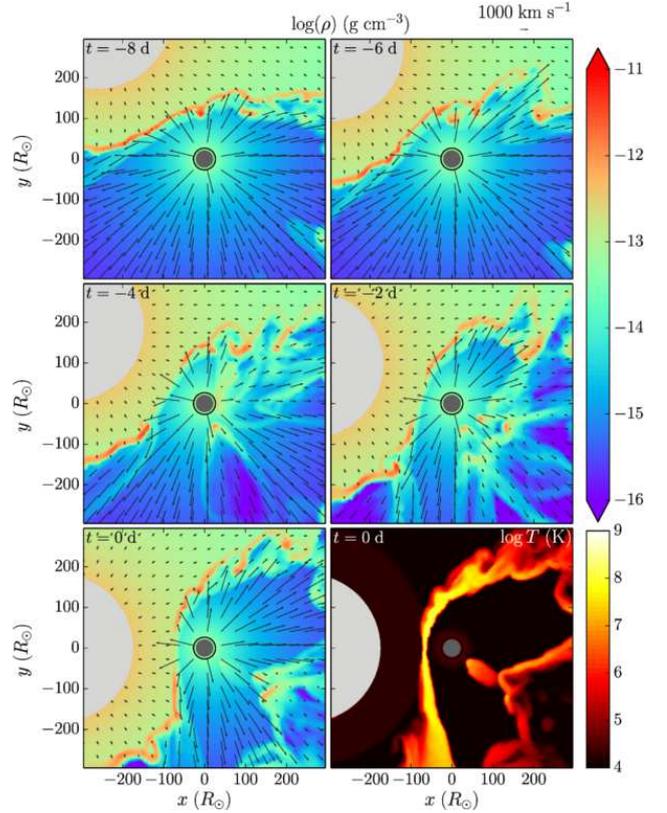}  
\caption{
Density maps with velocity vectors showing the density sliced in the orbital plane ($z=0$), for run C1, where we use the conventional mass model ($M_1=120 \rmModot$ and $M_2=30 \rmModot$).
The bottom right panel shows a temperature map.
The secondary is at the center, marked with a dark-gray circle while the primary, marked with a light-gray circle, orbits it counter-clockwise starting from the upper part of the figure at $t = -50$ days.
The simulation is performed in the secondary rest frame, and therefore the orbital velocity of the primary is subtracted from its wind velocity. Note that this effect is hardly seen in the figure due to the short arrows depicting the primary wind velocity.
Periastron occurs at $(x,y,z)=(-1.664 \au, 0, 0)$ and $t=0$.
Times are given with respect to periastron.
The secondary wind is being injected, at terminal velocity, between the secondary and the black circle around it.
We model gravity by two point masses at the locations of both stars.
Accretion starts $\simeq 5$ days before periastron when the dense clumps that formed in the post-shocked primary wind enter the injection region of the secondary wind, and lasts for only $\simeq 2$ days.
}
\label{fig:density_slices}
\end{figure}

In \citetalias{Kashi2017} we presented the results for a run with similar parameters to run C1 we show here.
In both runs we used the conventional mass model ($M_1=120 \rmModot$ and $M_2=30 \rmModot$). The differences are, as explained in section \ref{sec:simulation}, that we here model the flow without self-gravity but rather with two point-masses for the two stars, and that the accreted mass is removed from the simulation.
Figure~\ref{fig:density_slices} shows density maps for run C1 in the orbital plane ($z=0$), at different times of the simulation. 
Times are given with respect to periastron.
The secondary is at the center of the grid, and the primary orbits it from the upper part of the figure to the bottom-left until periastron, and then bottom-right.
At periastron the primary (light gray circle) is exactly to the left of the secondary (dark gray circle).
The secondary wind is being injected between secondary radius and the black circle at its terminal velocity.

Comparing run C1 in Figure~\ref{fig:density_slices} with figure~1 in \citetalias{Kashi2017}, it can be seen that accretion starts at the same time ($t \simeq -5$ days), but here the accretion time is shorter, lasting only for $\simeq 2$ days.
The reason for the difference is the removal of the accreted gas we enforce as part of approach 1. Taking the gas away allows the secondary wind to continue to blow without interruption.
Such an interruption existed in the run we presented in \citetalias{Kashi2017} by the untreated accreted gas that was left to accumulate around the secondary, and blocked some of the secondary wind, which in turn allowed more gas to be accreted.
%

\begin{figure}
\centering
\includegraphics[trim= 0.cm 0.cm 0.cm 0.cm,clip=true,width=0.99\columnwidth]{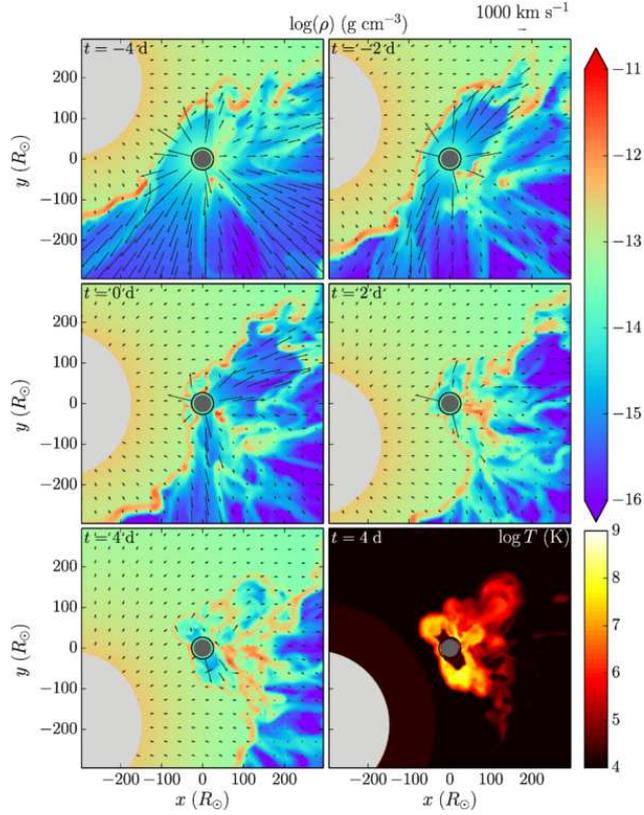}  
\caption{
Like Figure~\ref{fig:density_slices} but for run C2.
At times prior to 5 days before periastron there is (by definition) no difference, but then we lower the density of the secondary wind according to equation (\ref{eq:mdot_approach2}), therefore resulting in the accretion of more mass onto the secondary.
It can be seen that as of day $-4$ the colliding winds structure cease to exist. Then accretion occurs directly from the primary wind onto the secondary.
}
\label{fig:density_slices_run2}
\end{figure}

Figure~\ref{fig:density_slices_run2} shows the same time series of density maps for run C2 where we also use the conventional mass model but this time the second approach for secondary mass loss response to accretion.
The second approach is very artificial, in the sense that the mass loss rate of the secondary is being reduced regardless of the interaction details with the primary wind and accretion. It assumes that the accreted gas shuts down the secondary wind (more accurately, reducing its mass loss rate by a factor of 10).
We use this approach to obtain an upper limit for the accretion rate.
It can be seen that indeed as a result of reducing the mass loss rate of the secondary much more mass can reach the secondary and be accreted.
Over a duration of 70 days the secondary accreted $M_{\rm acc} \simeq 3.8 \times 10^{-6} \rmModot$.
We do not reinstate the original mass loss rate of the secondary wind during this time interval.
Had we done so the accretion would have most probably stopped at the time of reinstatement.
%

\begin{figure}
\centering
\includegraphics[trim= 0.cm 0.cm 0.cm 0.cm,clip=true,width=0.99\columnwidth]{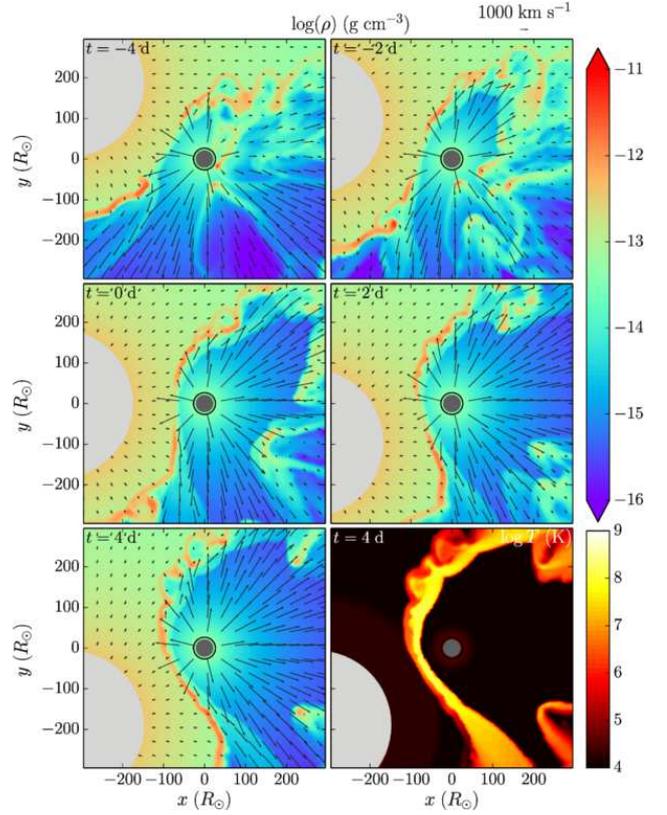}  
\caption{
Like Figure~\ref{fig:density_slices} but for run C3: conventional mass model with approach (3) for secondary wind response to accretion.
We find that there is not a very large difference from the results of run C1.
}
\label{fig:density_slices_run3}
\end{figure}

Figure~\ref{fig:density_slices_run3} shows the same time series of density maps for run C3 where we also used the conventional mass model but this time the third approach for secondary mass loss response to accretion.
The secondary wind does not retaliate much to the accreted mass, and there is no significant prolongation of the accretion phase compared to run C1.
%

\begin{figure}
\centering
\includegraphics[trim= 0.cm 0.cm 0.cm 0.cm,clip=true,width=0.99\columnwidth]{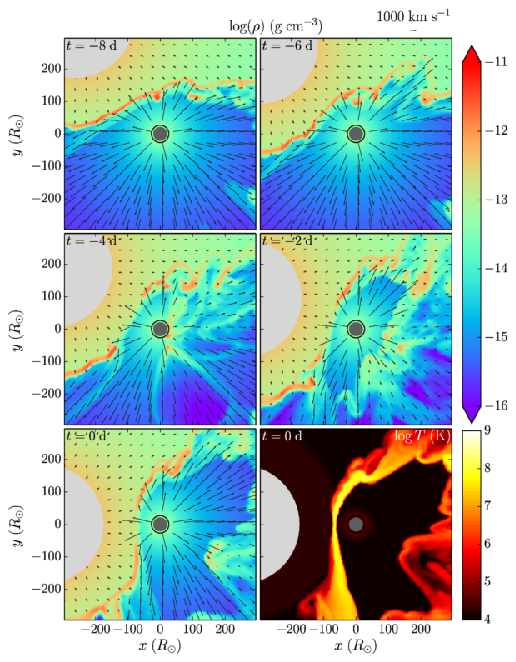}  
\caption{
Like Figure~\ref{fig:density_slices} but for run C4: conventional mass model with approach (4) for secondary wind response to accretion.
Mass accretion is larger than approaches 1 and 3, but much smaller than approach 4.
}
\label{fig:density_slices_run4}
\end{figure}

The simulation in which we adopted the last of our four approaches for the conventional mass model is shown in Figure~\ref{fig:density_slices_run4}.
We find that there is a considerably high accretion rate compared to approaches 1 and 3.
As the gas that reaches the wind ejection region is not removed from the simulation, it is able to penetrate deeper into the wind ejection region and spread on wider solid angles.
The result is regions in the secondary atmosphere that stop pushing wind as a result of accretion.
Consequently, the mass acretion rate is higher and the accreted mass accumulates to  $M_{\rm acc} \simeq 2 \times 10^{-7} \rmModot$ for the duration of accretion.
In this case accretion lasts for 2.5 days and later stops for about a month, after which there is a minor accretion of a clump lasting about 0.5 day.
We consider this a natural result of a blob accretion.
%

\begin{figure}
\centering
\includegraphics[trim= 0.0cm 0.0cm 0.5cm 0.5cm,clip=true,width=0.99\columnwidth]{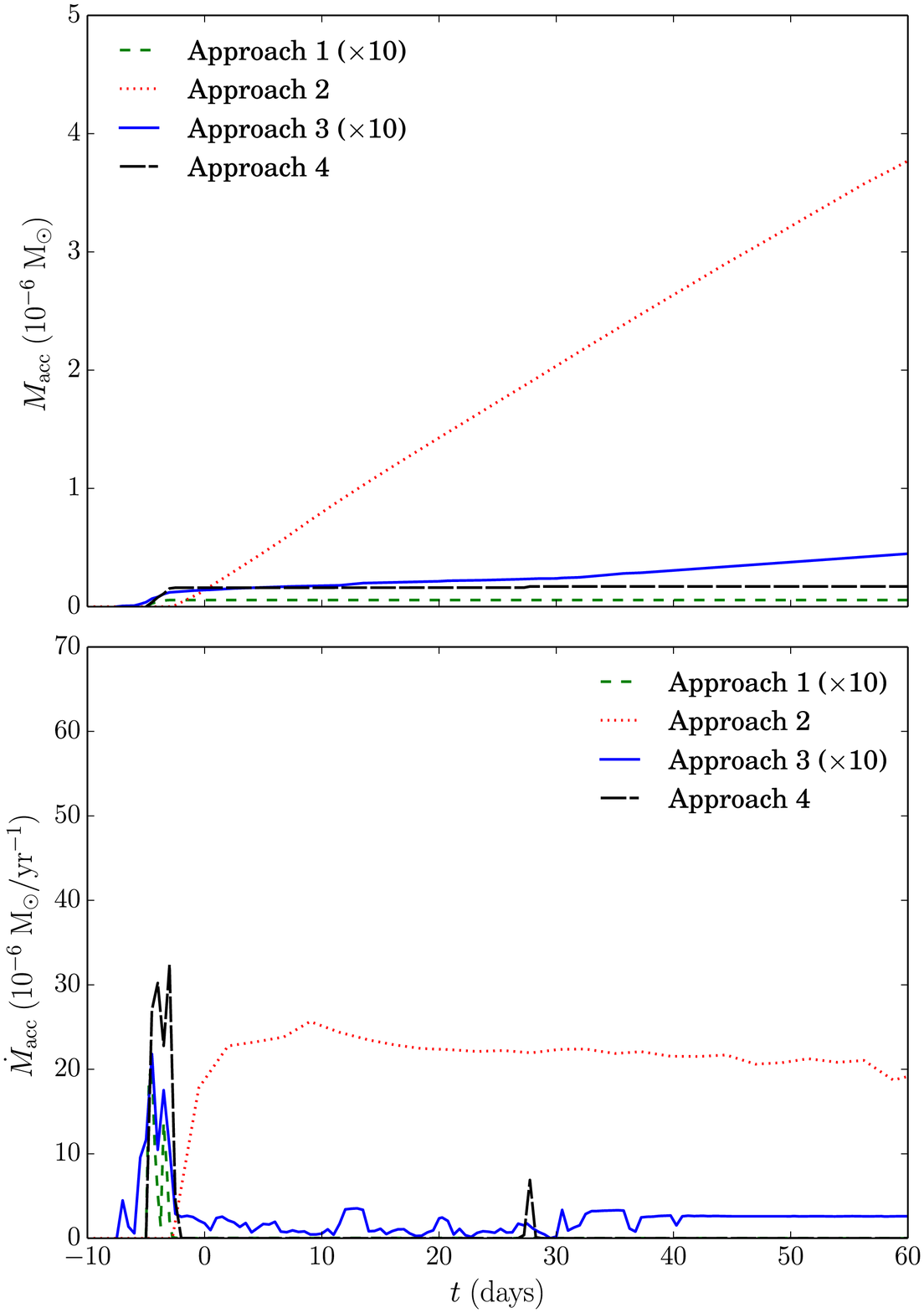}  
\caption{
The accreted mass (upper panel) and the accretion rate (lower panel) for runs C1--C4 (Table \ref{table:parameters}), where we simulated the conventional mass model ($M_1=120 \rmModot$ and $M_2=30 \rmModot$) with the four different approaches for secondary wind response to accretion.
Note that for clarity the values in both panels for Approach 1 and 3 have been multiplied by 10.
}
\label{fig:accretion12030}
\end{figure}

We compare the mass accretion rates of runs C1--C4 in Figure~\ref{fig:accretion12030}.
We can see that using approaches 1 and 3 gives very small accreted mass, while approach 4 yields larger amount.
Approach 2, as discussed above gives an indication for the upper limit, had the secondary wind been reduced to 10\% its mass loss rate for the duration of the event.
In all the runs for the conventional mass model runs (except run C2) the accreted mass is not large enough to account for observations according to the estimates in \cite{KashiSoker2009b}.

\begin{figure}
\centering
\includegraphics[trim= 0.cm 0.cm 0.cm 0.cm,clip=true,width=0.99\columnwidth]{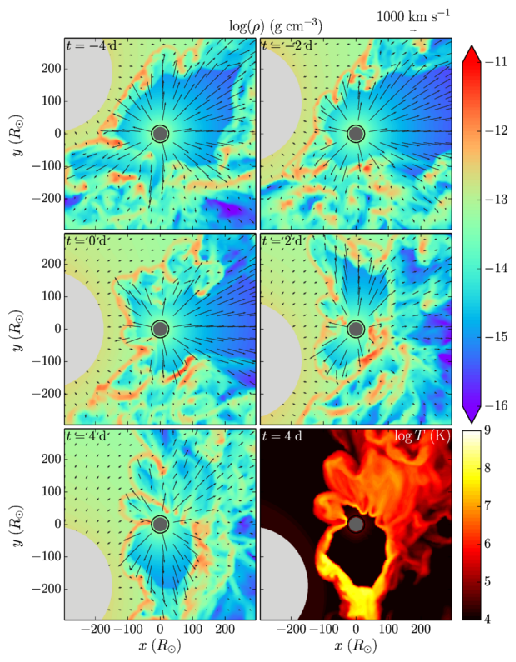}  
\caption{
Like Figure~\ref{fig:density_slices} but for run M4.
Here we use the high mass model, $M_1=170 \rmModot$ and $M_2=80 \rmModot$, while the orbital eccentricity and the mass loss rate of the primary are the same as in runs C1--C4.
}
\label{fig:density_slices_run_M4}
\end{figure}

We repeated the simulation for the high mass model (i.e., $M_1=170 \rmModot$ and $M_2=80 \rmModot$), keeping the orbital period and winds properties the same as in the previous simulation, and taking the same four approaches for the secondary wind response to accretion.
Figure \ref{fig:density_slices_run_M4} shows time series of run M4 where approach 4 for the response of the secondary to the accreted gas was used.
Already at $t=0$ the difference between runs M4 and C4 is evident.
On run M4 there is a vibrant accretion going on while on run C4 there is no memory of the weak accretion episode that took place only a few days earlier.
%

\begin{figure}
\centering
\includegraphics[trim= 0.0cm 0.0cm 0.0cm 0.0cm,clip=true,width=0.99\columnwidth]{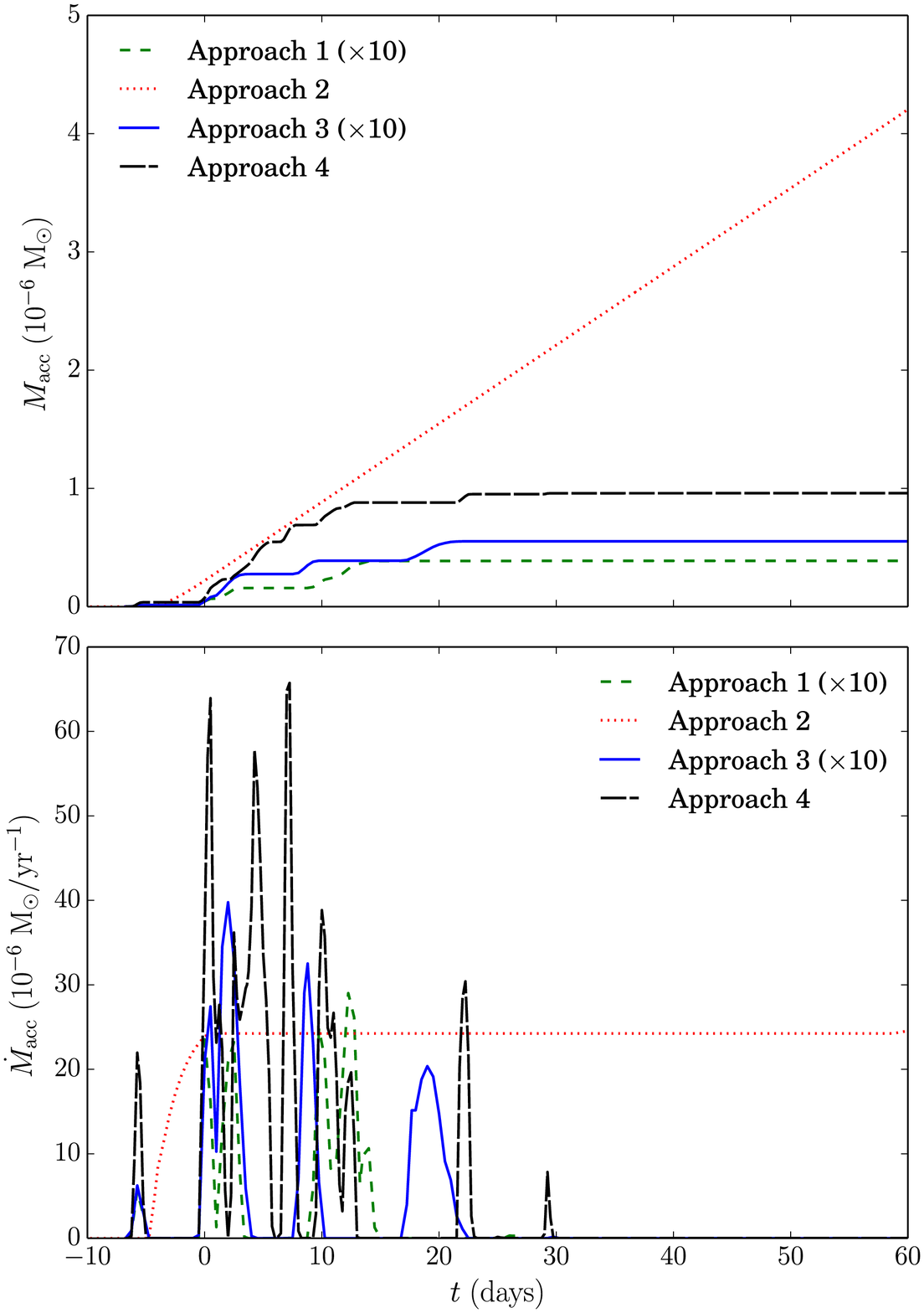}  
\caption{
Comparison of the accretion rates and accumulated accretion for runs M1--M4, where we simulated the high mass model ($M_1=170 \rmModot$ and $M_2=80 \rmModot$) with the four different approaches for secondary wind response to accretion.
Note that for clarity the values in both panels for Approach 1 and 3 have been multiplied by 10.
}
\label{fig:accretion17080}
\end{figure}

The results of the mass accretion rates and cumulated accreted mass for runs M1--M4 (high mass model) are shown in Figure~\ref{fig:accretion17080}.
The orbit for the high mass model has the same period and eccentricity, but the semi-major axis is larger, and consequently the periastron distance.
As the gravitational well of the secondary is deeper for the high mass model, the secondary can more easily accrete the filaments and clumps formed in the colliding winds structure, and therefore accretion starts $\simeq 7.5$ days before periastron ($\simeq 2.5$ days earlier than for the conventional mass model).
%

We find that the accreted mass for approaches 1 and 3 (runs M1 and M3, respectively) are still low.
For approach 2 (run M2) we see that the accretion rate reaches saturation at about $t\simeq 0$ days.
The saturation in mass accretion rate describes a situation where the secondary wind does not succeed to overcome the momentum of the accreted gas from the primary wind. The primary wind is
engulfing the star from almost all directions and the secondary wind is almost trapped. The secondary then accretes as much as it can.
The wind of the secondary escapes through gaps in the primary wind, creating bubbles of thin gas within the dense primary wind.
The rest of the gas of the primary wind, that cannot be accreted but is still gravitationally bound to the secondary, is accumulating around the secondary, waiting to be accreted.
Note that we do not get saturation for the conventional mass mode (run C2), as in this run the
secondary is able to push back up to 25$\%$ of the accreted wind. Namely, accretion in run C2 is weaker than in run M2.

Approach 4 (run M4) presents a different picture, with much larger mass accretion rate and accreted mass of $\simeq 1.0 \times 10^{-6} \rmModot$ over the duration of the simulation.
The stronger gravity of the secondary in the high mass model makes a significant difference, allowing much more mass to be accreted compared to the conventional mass model (run C4).

\begin{figure}
\centering
\includegraphics[trim= 0.8cm 0.1cm 1.5cm 1.2cm,clip=true,width=1.0\columnwidth]{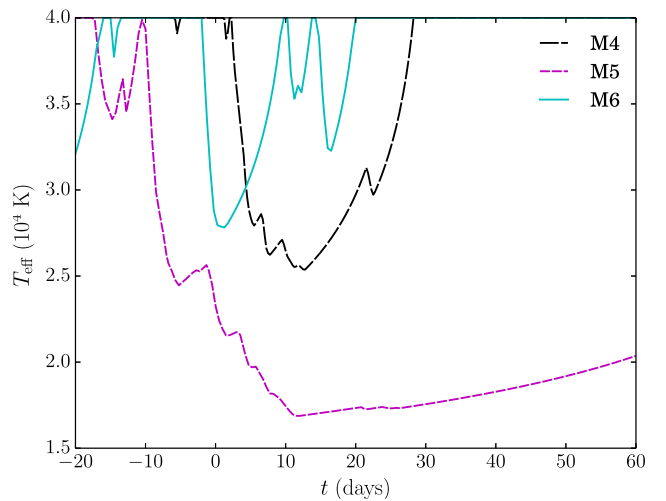}  
\caption{
The reduction in the effective temperature of the secondary, averaged over directions ($4\pi$). 
}
\label{fig:Tavg}
\end{figure}

As noted above, the accreted gas is expected to reduce the effective temperature of the secondary and results in emission lines of lower ionization states.
We post-process the simulations results to obtain the effective temperature of the secondary as a result of the obscuring gas.
For that, we first calculate the optical depth towards the star.
The optical depth is calculated from an outer radius $R_{\rm{out}}$ towards the center, stopping at $R_2$

\begin{equation}
\tau(t) = \int\limits_{R_{\rm{out}}}^{R_2} \!
-\kappa(r,t) \rho(r,t) \,\rm{d}r,
\label{eq:tau}
\end{equation}
where $\kappa(r,t)$ is the Rosseland opacity, which depends on time since the density and temperature at each position along the calculated path are time dependent. The density $\rho(r,t)$ is determined by the simulations results, taking into account the accreted mass and the mass loss of the secondary.
We then obtain the effective temperature assuming a grey photosphere, averaged over all directions
\begin{equation}
T^4_{\rm{eff}}(t) =   \frac{4}{3} \left(\tau(t) + \frac{2}{3} \right)^{-1} T^4(\tau=2/3),
\label{eq:Teff}
\end{equation}
where we take $T(\tau=2/3)=40\,000 \K$  to be the isotropic effective temperature of the undisturbed secondary,
though we note that recent analysis of UV lines may indicate higher temperature \citep{Davidsonetal2018}.
Figure~\ref{fig:Tavg} presents the effective temperature, showing the decrease as a result of accretion.
%

Observations of lines during the spectroscopic event indicate ionizing radiation from the secondary that is equivalent to that
of a star with an effective temperature of $\lesssim 25\,000 \K$.
From the decrease in effective temperature we conclude that the approach that best fits observation of the spectroscopic event is approach (4).
Namely, we find that our simulations
start and end accretion consistent with the observations without needing a prescription code that would intervene with the natural process.

Still, the temperature we obtained for run M4 is somewhat higher than indicated by observed lines during the spectroscopic event, suggesting that the amount of accreted gas should be higher than we obtained for the parameters we used in this run.
We therefore add more runs to see the effect of varying some of the parameters of the problem.
Since we are dealing with very expensive computation we cannot go through all parameters and all their ranges, and therefore we restrict ourselves to the important ones and to key simulations that will show trends in the amount of accreted gas and reduction of the effective temperature of the secondary.

Runs C5 and M5 are similar to C4 and M4, respectively, with the only change of increasing the primary mass loss rate to $\dot{M}_1= 10^{-3} \msyr$.
This value is supposed to resemble the older state of the primary mass loss rate, before its change of state (see section \ref{sec:intro} for details).
For run M5 we find that the change we did compared to run M4, in a factor of $5/3$ to $\dot{M}_1$,resulted an increase by a factor of $3.1$ in $M_{\rm acc}$, such that overall the accreted mass is $M_{\rm acc} \simeq 3.1 \times 10^{-6} \rmModot$.
The duration of accretion is about $20\%$ longer than run M4. It starts $\simeq 18$ days before periastron passage and lasts for $\simeq 45$ days.
This demonstrates the nonlinear dependency of accretion in the primary mass loss rate.
Figure~\ref{fig:Tavg} also shows the effective temperature for run M5, showing somewhat too strong drop in the effective temperature, and for longer duration than indicated by observations of the 2003.5 spectroscopic event.
We can therefore conclude that a mass loss rate larger than the one in run M4, but lower than the one taken in run M5, about $\dot{M}_1 \approx 8 \times 10^{-4} \msyr$ would best fit the properties inferred from observations.
Observations of later spectroscopic events that were shorter and had weaker variation in lines as a result of the secondary UV radiation better fit primary mass loss rate closer to the lower value of run M4.
It is interesting to note that \cite{Grohetal2012} found similar value, $8.5 \times 10^{-4} \msyr$ from 2D radiative transfer modeling of UV and optical spectra taken when the binary system was near apastron.
%
\begin{figure}
\centering
\includegraphics[trim= 0.0cm 0.0cm 0.0cm 0.0cm,clip=true,width=0.99\columnwidth]{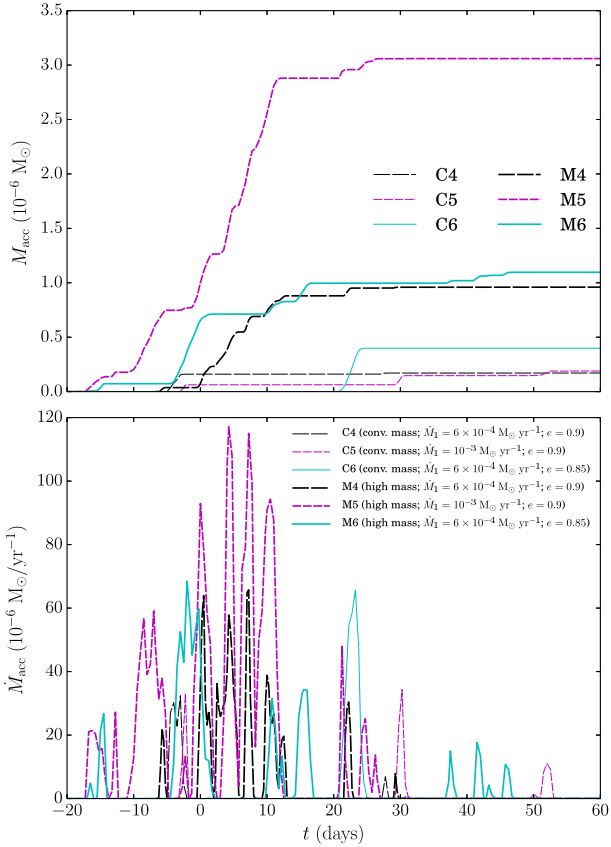}  
\caption{
The accreted mass (upper panel) and the accretion rate (lower panel) for runs C4--C6 and M4--M6 (see Table \ref{table:parameters}).
It can be seen that for the high mass model (M-runs) much more mass is accreted onto the secondary. The main reason is the stronger gravity of the secondary.
It is also very clear that stronger mass loss rate of the primary (runs C5 and M5) causes a large increase in the accreted mass.
The dependence on eccentricity is more complicates as lower eccentricity (runs C6 and M6) means larger periastron distance but also longer periastron passage.
These two effects can combine in different ways, as described in the text.
}
\label{fig:accretionC456M456}
\end{figure}

Another parameter we vary is the eccentricity, for which we also test the value $e=0.85$. This value was favored by \citealt{Davidsonetal2017}, who mentioned that it gives the smallest possible separation distance at the critical time when the spectroscopic event begins.
It would therefore be expected that $e=0.85$ would produce earlier accretion compared to $e=0.9$, even though the periastron distance is $50\%$ larger for the smaller eccentricity.
Runs C6 and M6 (Figure~\ref{fig:accretionC456M456}) show that for $e=0.85$ the accretion duration is longer.
This confirms the claims of \citealt{Davidsonetal2017} since the time it takes for the secondary to undergo periastron passage is longer.
Run M6 also shows early accretion exactly as expected by \citealt{Davidsonetal2017}.
In run C6 we have not seen this behaviour, and the reason is that the larger periastron distance and smaller secondary mass combined to reduce the gravitational attraction of the secondary and therefore early accretion could not occur.
Both runs C6 and M6 produced larger mass accretion than their counterparts with $e=0.9$, runs C4 and M4, respectively.

The last effect we test, in a preliminary way, is the acceleration of the secondary wind.
In run M4WA we take the parameters as in run M4, but instead of ejecting the wind at terminal speed, we accelerate the secondary wind using a $\beta$ profile motivated by the traditional CAK model \citep{Castoretal1975}. We take $\beta=1/2$, a value considered to be appropriate for O-stars (e.g., \citealt{Vinketal2011}), for which the wind is accelerated according to an inverse-$r^2$ law, with radial acceleration
\begin{equation}
a_2(r_2)=\frac{v_{2,\rm{inf}}^2}{2R_2} \left(\frac{R_2}{r_2}\right)^2,
\label{eq:beta_half_acc}
\end{equation}
where $r_2$ is the distance from the center of the secondary and $v_{2,\rm{inf}}=3\,000 \kms$ is the terminal velocity of the secondary wind.
We apply the acceleration only to material that has a velocity vector in the radial direction. If any cell has a velocity vector deviating from the radial direction, we treat it as affected by incoming gas, and stop accelerating it.

\begin{figure}
\centering
\includegraphics[trim= 0.0cm 0.0cm 0.0cm 0.0cm,clip=true,width=0.99\columnwidth]{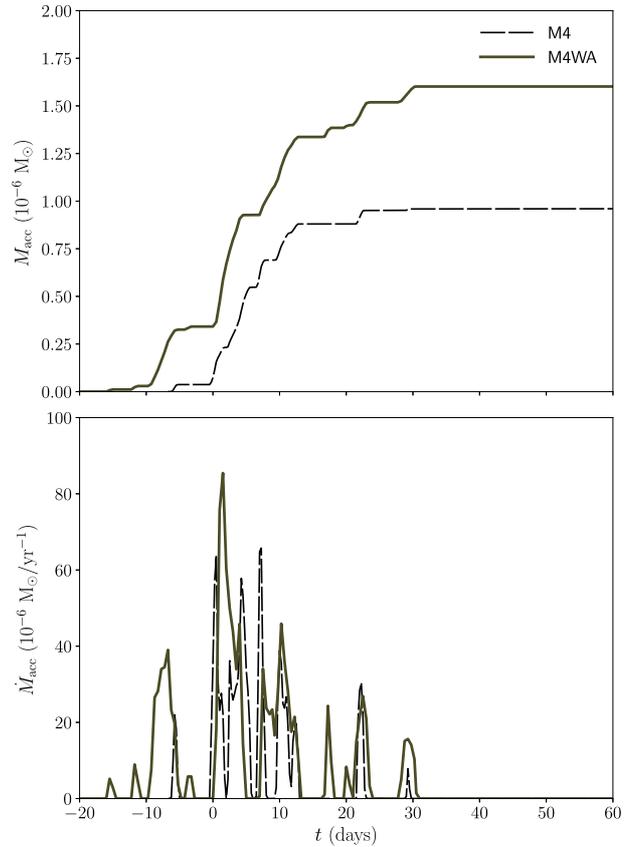}  
\caption{
The accreted mass (upper panel) and the accretion rate (lower panel) for run M4 for which the wind of the secondary is ejected at terminal velocity, and run M4WA for which the wind of the secondary is accelerated according to a $\beta=1/2$ profile.
Accretion for the accelerated wind case is larger by $\simeq 60 \%$, and the duration is longer by $\simeq 26 \%$.
}
\label{fig:accretionM4M4WA}
\end{figure}

Figure \ref{fig:accretionM4M4WA} compares the result of run M4 and M4WA. We caa see that the accretion rate obtained when taking the wind acceleration into account is somewhat larger, $M_{\rm acc} \simeq 1.6 \times 10^{-6} \rmModot$ instead of $\simeq 1.0 \times 10^{-6} \rmModot$.
This result is expected since the secondary wind posing accretion is essentially less energetic when accelerated, rather than launched at terminal velocity.
%

\section{SUMMARY AND DISCUSSION}
\label{sec:summary}

We perform detailed 3D numerical simulations of the \etc colliding winds system close to periastron passage and derive the accretion rates onto the secondary star.
The colliding wind region is prone to instabilities that lead to a non-linear formation of clumps and filaments that are accreted onto the secondary star.
The accreted mass disturbs the secondary wind and weakens the mass loss, enabling more mass to be accreted.
The accretion finally stops in the post-periastron phase when the stars get further from each other and consequently the density of the primary wind at the vicinity of the secondary decreases.

Simulating accretion onto stars is a complicated task.
The code used for this assignment needs to be able to handle both the hydrodynamics of the flow as well as the interaction of the photons emitted from the star with the gas.
The \texttt{FLASH} code we use incorporates a radiation transfer unit which treats the photon-gas interaction, so the momentum of the accreted gas is being changed appropriately along its trajectory.
In our simulations we included only the energy from the hot gas itself, which has a small effect on the gas that already has relatively high velocity. The acceleration from the photon emitted from the stars was not included directly but rather implied by the initial velocities given to the winds (and in one case it was implied in their acceleration in run M4WA where the beta-profile was used).
The FLASH code can only treat the solution of the radiative transfer equation in the absence of scattering, namely it offers only a formal solution of the radiative
transfer equation and not the full self consistent scattering solution. The effect of scattering may be addressed using different methods, such as dedicated radiation transfer codes that use monte-carlo approach.


We here focus on the way the secondary wind would respond to the high accretion rate, and for that purpose suggest the four approaches discussed in section \ref{sec:simulation}: approaching gas removal, exponentially reduced mass loss, accretion dependent mass loss and no intervention.

We find that accretion is obtained for both the conventional mass model $(M_1,M_2)=(120 \rmModot, 30 \rmModot)$ and the high mass model $(M_1,M_2)=(170 \rmModot, 80 \rmModot)$. 
For the high mass model the stronger secondary gravity attracts the clumps and we get higher accretion rates and longer accretion period.

Obviously our simulations are not full radiation-transfer simulations, and as such do not provide complete details regarding the ionization structure. However they are sufficient to show the reduction in the secondary effective temperature within the assumption of high optical depth.
We show that for the runs where accretion is substantial, $M_{\rm acc} \gtrsim 10^{-6} \msyr$, the effective temperature of the secondary drops as a result of the ambient gas.
Consequently fewer ionizing photons are emitted from the secondary, which is the major ionizing source of the binary system despite its lower luminosity.
Therefore the ionization structure changes for the duration of the event.
This confirms the basic idea of the accretion model that the obscuration of the ionizing photons of the secondary is the cause for variations in lines during the spectroscopic event \citep{Soker2001,Soker2005a,Soker2005b,Soker2007,Akashietal2006}.

One important parameter we studied is the mass loss rate of the primary.
We used values within the range of values explored in the literature (as discussed in \citetalias{Kashi2017}).
Our simulations demonstrated that the mass loss rate of the primary affects the accretion rate of the secondary in non-linear way.
Our results suggest that at least for the 2003.5 and 2009 periastron passages the mass loss rate of the primary was $\dot{M}_1 \approx 8 \times 10^{-4} \msyr$, similar to the value obtained by \cite{Grohetal2012} from observations.
The 2014.6 spectroscopic may have implied a further decrease in the primary mass loss rate, which was claimed to be an ongoing trend \citep{Mehneretal2015}.
A similar decrease in the primary mass loss rate was obtained by simulations of recovery from giant eruptions \citep{Kashietal2016}.
Our simulations partially support the conclusions of \citep{Mehneretal2015} who suggested that if the mass loss rate of the primary continues to decrease we will have very weak spectroscopic events in the future or there may be none. We indeed find strong dependency between the accreted mass and the mass loss rate of the primary, and it is clear that if the mass loss rate of the primary is lowered by a factor of a few the accretion can stop.
Moreover, \cite{Mehneretal2015} claimed that in the 2014.6 event the primary mass loss rate has fallen low enough so that full accretion have not occurred, as opposed to previous events.
Our simulations are unable to confirm or refute this conclusion due to the uncertainty in many parameters.

In our study we counted material as accreted only if it actually reached the secondary. 
It is noteworthy to mention that there are other examples in the literature for accretion criteria, such as
reaching the outer edge of the wind ejection zone (e.g. \citealt{Akashietal2013}).
One very relaxing criteria is the one brought by \cite{deVal-Borroetal2009}, who preformed simulations of accretion in symbiotic systems.
When modeling Bondi-Hoyle-Lyttleton accretion \citep{HoyleLyttleton1939,BondiHoyle1944} they removed some of the gas in the vicinity of the star and added it to the accreting star.
For that they used a criterion for the accretion radius to be $R_{\rm acc} = 0.1 R_H = 0.1 r (M_2/3M_1)^{1/3}$, where $R_H$ is the Hill radius of the accreting star and $r$ is the binary separation.
In our case adopting such an expression for the conventional mass model (the high mass model) would give $R_{\rm acc} \simeq 15.6 \rmRodot$ ($\simeq 22.9 \rmRodot$) at periastron, and larger accretion radii before and after periastron time.
While for the conventional mass model $R_{\rm acc} < R_2$, adopting prescription of \cite{deVal-Borroetal2009} for the accretion radius in the case of the high mass model of \etc would have given a significantly wider accretion radius, and would consequently increase the accretion rate considerably.

As expected, we can see that the accretion rate obtained when taking the secondary wind acceleration into account was higher.
It is not very difficult to come up with improved approaches for how the secondary wind would react to accretion.
For example, our third approach neglects possible rotation of the secondary that would make the change in the mass loss rate to be closer to latitude dependent rather than direction dependent.
Even though more sophisticated approach can be used, we find the ones we use here to give accretion rates that match earlier estimates \citep{KashiSoker2009b} based on observations of the duration of the spectroscopic event.
We therefore leave the development of higher order approaches to a further study.
These will also take into account the full effects of wind acceleration, for which we show here preliminary results.

Accelerating the stellar winds will lead to two competing effects that might affect the accretion onto the secondary. The pre-shock velocity will be lower, which will reduce the penetration of the clumps and their ability to reach the secondary’s surface, while the pre-shock density will also be higher, thereby undergoing more radiative cooling and creating denser clumps that will have a higher chance of reaching the secondary star. As such, the incorporation of wind acceleration of both stars will be a key component of future work.
The lower accretion rates compared to \cite{KashiSoker2009b}, where the accelaration was taken into account, may suggest that the acceleration of the wind has a role in the details of accretion and the response of the secondary to accretion.

An additional aspect is the way the accreting star would respond to the accreted gas which settles onto its envelope, with momentum in the opposite direction to its blowing wind, angular momentum, and different composition.
A complete treatment requires involving a stellar evolution code and adding the accreted mass to the secondary star and obtaining the properties of the wind as a result.
It may also be required to iterate between the hydrodynamical and stellar evolution code in order to get a consistent solution.
However even the most modern stellar evolution codes only use formulated (or semi-empirical) prescriptions for mass loss rates (e.g. \citealt{KudritzkiPuls2000,Vinketal2001,Vinketal2011,Pulsetal2008,Vink2015}).
It is therefore not clear that the exercise suggested above will produce better results than the treatment we incorporate here.

As the parameter space is large and not tightly constrained, there is no point at this time to fine tune the parameters in our simulations.
The main point is that some of the secondary wind response approaches we explored match the properties inferred from observations, and better support the high mass model for \etc.
In a future work we intend to explore in more details directional effects of the accreted gas and quantitatively study the angular momentum of the accreted gas and how it effects the binary system at times of spectroscopic events.

\section*{Acknowledgements}
I appreciate very helpful comments from Noam Soker and an anonymous referee.
This work used the Extreme Science and Engineering Discovery Environment (XSEDE) TACC/Stampede2 at the service-provider through allocation TG-AST150018.
This work was supported by the Cy-Tera Project, which is co-funded by the European Regional Development Fund and the Republic of Cyprus through the Research Promotion Foundation.
This research was enabled in part by support provided by Compute Canada (\url{www.computecanada.ca}), thanks to the sponsorship of A. Skorek.

\label{lastpage}
\end{document}